# CLICK SPAM PREVENTION MODEL FOR ON-LINE ADVERTISEMENT

Nicola Zingirian and Michele Benini

Department of Information Engineering, University of Padova, Italy

nicola.zingirian@unipd.it
michele.benini@studenti.unipd.it

## ABSTRACT

*This paper shows a vulnerability of the pay-per-click accounting of Google Ads and proposes a statistical tradeoff-based approach to manage this vulnerability. The result of this paper is a model to calculate the overhead cost per click necessary to protect the subscribers and a simple algorithm to implement this protection. Simulations validate the correctness of the model and the economical applicability.*

## 1. INTRODUCTION

The pay-per-click accounting method adopted by Google Ads service for online advertising services [1][2] enables the Advertising Provider (AdP, e.g., Google) to automatically charge the Advertised Subscribers (AdS) for each single advertised page access (called "click") requested by each web user. Differently from previous online pay-per-click methods, this method does not need AdP and AdS to agree on respective web access logs and "referrer headers" [3], as a consequence the cost of subscriber's signup and charging processes are extremely reduced. This method allows a single AdP to manage millions pay-per-click contracts, thus making the pay-per-click advertising a mass service, as it appears today.

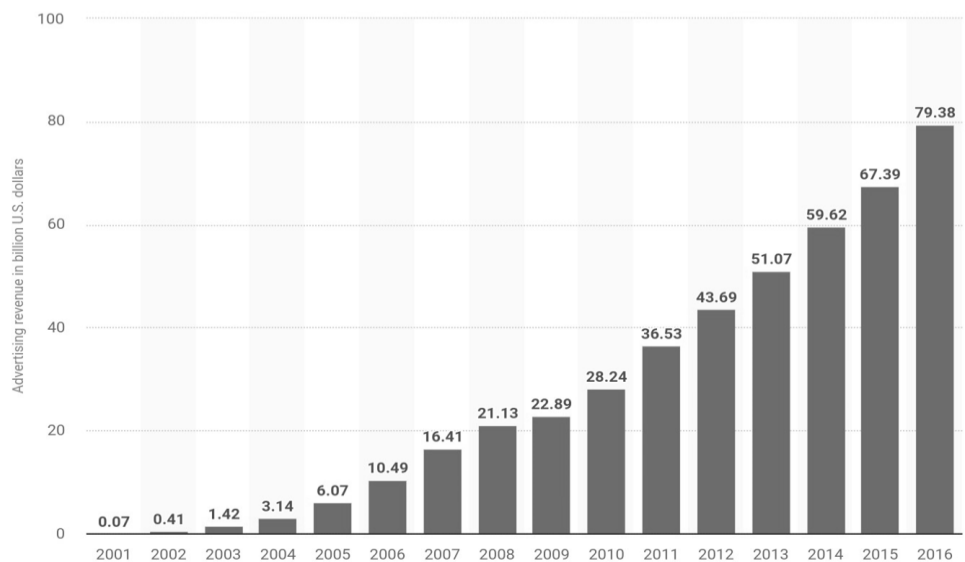

Figure 1 - Google Ads Revenue [10]

Being the click count the key figure upon which such a mass business calculates billion dollars revenue (see Figure 1), the paper presents a contribution to the relevant topic of evaluating the reliability of such a figure. In particular, the paper focuses on the robustness against malicious web clients who might spam the click counts, typically to make the click count rapidly exhaust the daily budget, thus eliminating AdS competitors from the advertising network within the first hours of each accounted day [4]

While big effort has been spent to setup heuristics to detect distributed click spam [5] [6][7] [8], we show in this paper that even a centralized attack works. In particular we shows how a simple malicious web user agent, using one IP address only, can make the click count increase for a given AdS, even if the charged clicks do not correspond to any real advertisement (Section 2).

To protect the AdS from this type of attack, we present a detailed statistical model of the trade-off between the security benefits obtained and the revenue losses caused by discarding potentially spammed clicks (Section 3).

Upon this model, we formulate a simple algorithm to control the trade-off (Section 4) and we validate it through simulation (Section 5).

Some remarks on applications and future work conclude the paper (Section 6).

## 2. ATTACK

### 2.1 METHODOLOGY

The attack presented in this paper, as depicted in Figure 2, is directed to the "fairness principle" of the AdP who charges one click only to the AdS, even when one user agent accesses the same advertised web page more than once. This principle considers that only the first click corresponds to a real advertisement, whilst the other repeated clicks do not provide any benefit in terms of advertised contents.

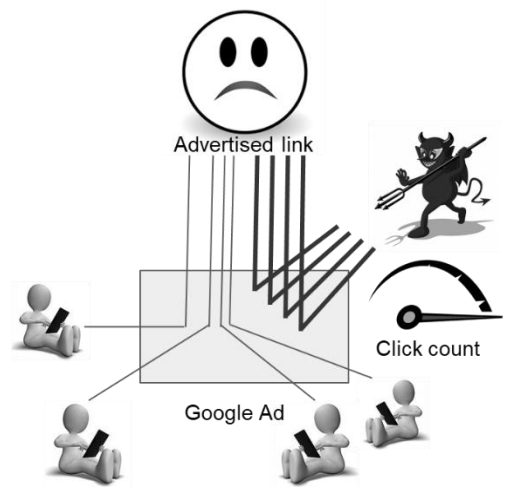

Figure 2 – Attack method

The vulnerability to this attack depends on the fact that no deterministic and secure rules to implement this principle are applicable on the AdP server side, to determine whether two clicks are originated by the same user agent or not.

According to the experiments done, the approach followed by Google algorithm adopts the heuristic to consider *n* clicks as originated by the same user agents according two conditions

(a) if the user agent "cookie" header values is the same in the *n* HTTP requests corresponding to the *n* clicks OR

(b) if the *n* clicks are originated by the same IP address and also *n* DNS queries are originated from the same IP address to resolve the AP Server name, just before each of the *n* HTTP request corresponding to the clicks otherwise the *n* clicks are both accounted, as if they were originated by two different user agents.

Consequently, the attack works as follows. A malicious HTTP client, whose pseudocode is available in Figure 3, performs many HTTP requests to the same link advertised by the AdP server, simply resetting both the cookie header value and the client's system DNS cache before sending each HTTP request. The client sends the request from the same source IP address.

Figure 4 and Figure 5 show the clicks accounted shown by Google Ads dashboard, before and after our experiments, respectively.

```
HTTP header  = "
      Host: // is set accordingly to the HTTP server name
      Accept:text/html,application/xhtml+xml,application/xml;q=0.9,image/webp,*/*;q=0.8
      Upgrade-Insecure-Requests: 1
      Accept-Language: it-IT,it;q=0.8,en-US;q=0.6,en;q=0.4,es;q=0.2,pl;q=0.2
      Connection: keep-alive
      User-Agent: //is set with a random choice of user-agent
      Accept-Encoding: */* "
repeat N times
      clean_DNS_cache();
      HTTP_Send ("GET www.google.it/search?q=<QUERY> HTTP/1.1" + HTTP Header);//No Cookies
      HTTP_Response = HTTP_receive()
      Response body = HTTP_parse(ENTITY_BODY, HTTP_Response)
      Target link = search(TARGET, Response body) // Attack Target
      MyCookie = http_parse(COOKIES, HTTP Response)
      HTTP_send ("GET target link HTTP/1.1" + HTTP Header + "Cookie = MyCookie")
      HTTP_Response = HTTP_receive()
      Status Code = http_parse(STATUS_CODE,HTTP Response)
      while (Status Code == 300) // Redirect
             Location = http_parse (LOCATION,HTTP Response)
                    HTTP_Send "GET Location HTTP/1.1" + HTTP Header + "Cookie = MyCookie"
                    Extract Status Code from HTTP Response
             end while
end repeat
```

Figure 3 Malicious HTTP client pseudocode.

Figure 4. Google AdWords daily accounted clicks before the attack

Figure 5. Google AdWords daily accounted clicks after the attack

In case of this sample attack, the AdP has detected only about 50% of false clicks, while the others are normally accounted, as shown in Figure 3.

The reader could wonder why Google heuristic does not consider the IP address as a safe information to determine that two clicks originate from the same user agent. The answer is that most IP networks over the Internet are IPV4 network adopting the Network Address Translation (NAT) [9], so that, potentially, two or even more hosts, each running a different user agent, can send distinct HTTP requests from the same IP address to the same AdP server. The AdP server might see, in that case, different clicks of really different users coming from the same IP address, as shown in Figure 6.

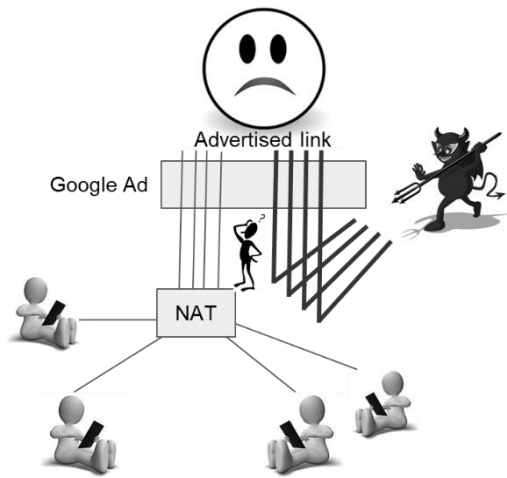

Figure 6 – Problems in IP address-based attack discrimination in NAT networks

We observe that disregarding the source IP address is a clearly safe heuristic for the AdP, unfortunately it is unsecure for the AdS as our experiments put in evidence.

Our investigation bases on the probability, for each NAT network, that two independent clicks, although coming from the same IP address, originate from two or more distinct user agents. This probability will drive the algorithm to decide whether counting such clicks or not.

## 2.1 RESULTS

Table 1. Attack result table

|  | After attack (started at 12.00 am) | | | End of the day | | |
|---|---|---|---|---|---|---|
| Date | Simulated click | Counted click | Discarded click | Counted click | Discarded click | Cost |
| 13/01/18 | 10 | 10 | 4 | 14 | 5 | 2,66* |
| 14/01/18 | 10 | 8 | 2 | 14 | 4 | 1,71 |
| 15/01/18 | 15 | 12 | 3 | 12 | 21 | 1,01 |
| 17/01/18 | 20 | 10 | 12 | 29 | 17 | 2,98* |

*THE DAILY BUDGET HAS BEEN EXCEEDED

Table 1 shows the results of the most recent experimental test performed just before the submission of this paper.

Every attack started at 12.00 a.m. when the click accounting resets. Between consecutive clicks the client waits a variable random time interval of 20 minutes average, as we noticed that more frequent or regular time intervals often reduce the attack efficiency.

The click counters figures have been extracted by report of the Google Ads dashboard one hour after the end of the attack execution, i.e., after Google's heuristics has discarded possible spammed clicks. The results show that the impact of spamming for the attacked AdS is huge, compared with the click counted at the end of the day reported in the right section of Table 1.

When the click accounting approaches the daily budget, Google AdWords stops advertising the contents, as in the cases reported in first and last table rows, so that the competitors of the attacked AdS can be strongly motivated in this type of attacks to increase their own visibility.

## 3. TRADE-OFF MODEL

The model analyzes the clicks coming from a NAT address space using the following variables

- a time interval $T$, called statistics window, within which two clicks are counted only once if they are activated by the same user and have the same target. If the time distance between two following clicks is more than time T they are counted twice although directed by the same user to the same advertised resource.
- an integer number $A$, corresponding to the cardinality of the NAT address pool.
- an integer number $C$ corresponding to the number of clicks coming from any address in the NAT pool and directed to the same AS resource

The average number $N(A,C)$ of "repeated clicks", i.e., the clicks directed to the same resource and coming from the same IP address but from different users over a NAT pool having cardinality A, provided that C clicks in total are coming from that NAT pool.

The case of "no malicious users involved" corresponds to supposing that each user performs only one click, so that repeated clicks are originated by two different users using the same IP address.

We define loss factor

$$L(A,C) = N(A,C) \times A / C$$

as the average percentage of real clicks that are potentially lost (i.e. not accounted) if all repeated clicks (2,3,4...), coming from the same address are systematically ignored. The average number of clicks N exceeding the single click is multiplied by the number A of addresses available, to obtain the average click over all the NATted network, and is divided by C to calculate the percentage over all clicks. This percentage corresponds to the loss rate of revenue that is paid by the AdP to protect the AdS.

If L is below a fixed threshold e.g. 1% , the heuristic decides to ignore all repeated click.

To calculate N(A,C) we consider the clicks as uniformly random independent events falling in a 1-d continuous space that splits into A equal segments, each representing the subspace of probability that a click is originated by the address represented by that segment, as shown in Figure 7. We assume that there is no correlation between IP address and user interest for a specific content, as there is no reason to suppose any correlation. This corresponds a Poisson Distribution where λ = C/A.

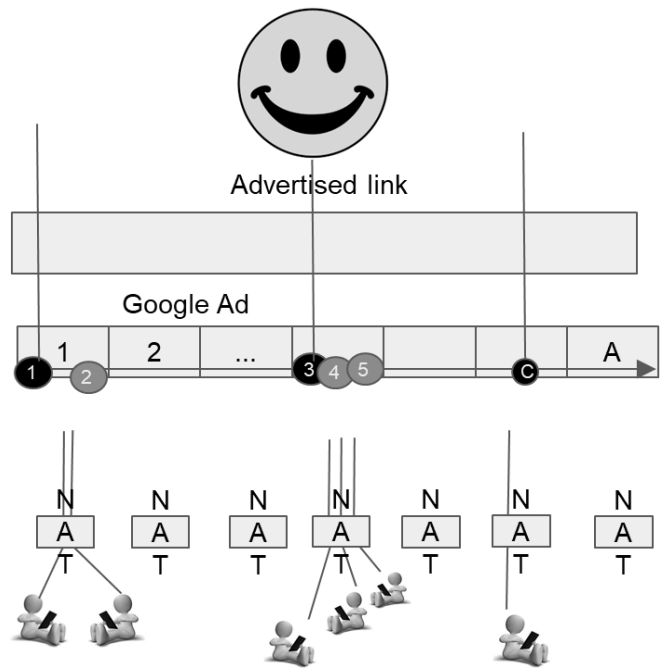

Figure 7 –Scheme of the Poisson process mapped to the click spam model

Considering that the average number of repeated events falling in the same interval is

$$\sum_{c=2}^{C}(c-1)\cdot P(clicks=c) = \sum_{c=2}^{C}(c-1)\cdot \frac{\lambda^c e^{-\lambda}}{c!}$$

Then

$$N(A,C) = \sum_{c=2}^{C}(c-1)\cdot \frac{(C/A)^c\, e^{-C/A}}{c!} = \frac{C}{A} + e^{-C/A} - 1$$

And consequently

$$L(A,C) = N(A,C)\cdot \frac{A}{C}$$

As shown in the Appendix, it is worth noticing that if A >> C then

$$N(A,C) < \frac{1}{2}\, C^2/A^2$$

As a consequence

$$L(A,C) < \frac{1}{2}\, C/A$$

This approximation, being a second-order Taylor polynomial approximation, is very precise for typical C/A values for large networks, e.g. If C/A = $10^{-3}$ then $\left|L(A,C) - \frac{1}{2}\, C/A\right| < 10^{-6}$.

## 4. ALGORITHM

The proposed algorithm is based on the following entities:

A table, called the *status table* having the following fields:
- <u>dest</u> : the destination URL of the Click
- <u>source</u>: the source IP address (possibly NATted)
- <u>net</u>: the id of the smallest network range registered in whoid DB
- <u>time</u>: timestamp (epoch)

An object, called click, having the properties of each click received, i.e.,
- click.dest : the destination URL of the Click
- click.source: the source IP address (possibly NATted)
- click.time: the click message receiving timestamp (epoch)

The actions done after each click is to count the click as valid incrementing the click counter through the function increment_counter() or to reject the click through the function "discard". The algorithm discards the clicks as long as the average statistical number of repeated clicks is below a threshold.

```
click_handler (click) {
      NET = lookup_net_by_ip(click.source); // from whois DB
      A = lookup_net_size(NET); // smallest whois DB net range
      delete from status_table where time < click.time - T; //removes oldest history
      C = select count(*) from status_table where dest = click.dest and net = NET; // calculates C
      if (select count(*) from status_table  where dest = click.dest and source = click.source > 0)
      { // if more than one click from that source.
      if (0.5 * C / A < threshold )  {
                discard(click);
                return;
                }
          }
          increment_counter (click.dest); // click is accounted for invoicing
      insert (click.source, click.dest, NET, click.time) into status_table;
}
```

Figure 8. Pseudocode of click selection handler

## 5. SIMULATION

We considered in our simulation the Italian provider Vodafone IT having 28.870.000 subscribers and 5.538.048 IP addresses registered. Supposing an advertisement having a huge impact e.g., the percentage of 0.1% over the whole population clicks the same advertised link then C= 28.870 and C/A = 5.21 $\cdot$ $10^{-3}$ .

According to the model presented in Section 4, the click loss ratio is ½ C/A = 2.6 $\cdot$ $10^{-3}$ if repeated clicks for each IP address are never accounted i.e. it is less than 0.26 %.

A simulation has been carried out to confirm the model.

The simulation programs:

- distributes N users randomly over the all IP address
- selects randomly N/1000 users who clicks the advertised link
- Counts how many clicks have been originated by each IP address

- Yields the number of IP addresses that originated 2 or more clicks divided by the total number of addresses

The simulation has executed 100.000 times and yielded the average value of $2.5368 \times 10^{-3}$.

The difference between model and simulation is:

$$|2.6065 \times 10^{-3} - 2.5368 \times 10^{-3}| = 6.9652 \times 10^{-5}$$

As a result, the impact of the loss is very low, even in case in which the number of clicks $C$ is very high, and the discrepancy between the model and the simulation is negligible. The cost of the protection in the Vodafone IT scenario would correspond to less than 0.26% discount in click accounting.

## 6. CONCLUSION

The tradeoff protection model presented in this paper allows the AdP precisely assigning a part of investment as an insurance to protect the customers from a very simple attack. The threshold used in the algorithm exactly corresponds to the percentage of turnover loss that can be decided by the AdP. Accepting this loss, the AdP protects the customer from click spam coming from a single IP address.

This model is particularly suitable for the large mass of small AdS who receive a few clicks per day for which a single repeated attack completely vanishes their investments. In that case, the customer should be asked to buy an insurance fee increasing the click fee by e.g., 0,5% to 3% of the click price to get the protection against this attack or the AdP can offer this feature as a quality of service parameter.

The time interval T should be large enough to collect enough statistics on $C$, and small enough to keep the memory of the clicks not too large to count the accesses of the same users who click again the same resource after long time.

This aspect is the key of evolution of the presented algorithm, as the standard deviation of the repeated clicks will be also considered to calculate the loss more precisely, possibly postponing the calculation when the mean is considered enough significant.

## 7. APPENDIX

Being

$$N(A, C) = \sum_{c=2}^{C} (c-1) \cdot \frac{(C/A)^c \, e^{-C/A}}{c!}$$

By replacing $\lambda = \frac{C}{A}$, we obtain

$$N(\lambda) = \sum_{c=2}^{C} (c-1) \frac{\lambda^c e^{-\lambda}}{c!}$$

that can split into two the following parts:

$$N(\lambda) = \sum_{c=2}^{C} c \frac{\lambda^c e^{-\lambda}}{c!} - \sum_{c=2}^{C} 1 \cdot \frac{\lambda^c e^{-\lambda}}{c!}$$

The first part is

$$\sum_{c=2}^{C} c \frac{\lambda^c e^{-\lambda}}{c!} = \sum_{c=0}^{C} c \frac{\lambda^c e^{-\lambda}}{c!} - \lambda e^{-\lambda} - 0 =$$

$$= \lambda - \lambda e^{-\lambda}$$

because $\sum_{c=0}^{C} c \frac{\lambda^c e^{-\lambda}}{c!}$ corresponds by definition to the Poisson mean, i.e., $\lambda$.

The second part

$$\sum_{c=2}^{C} \frac{\lambda^c e^{-\lambda}}{c!} = \sum_{c=0}^{C} \frac{\lambda^c e^{-\lambda}}{c!} - (e^{-\lambda} + \lambda e^{-\lambda})$$

$$= 1 - (e^{-\lambda} + \lambda e^{-\lambda})$$

Because $\sum_{c=0}^{C} \frac{\lambda^c e^{-\lambda}}{c!} = 1$ as it is the sum of the whole Poisson distribution.

Re-composing the two parts, we obtain

$$N(\lambda) = \lambda - \lambda e^{-\lambda} - 1 + (e^{-\lambda} + \lambda e^{-\lambda}) =$$

$$= \lambda + e^{-\lambda} - 1$$

Expanding N($\lambda$) the Taylor's series of $N(\lambda)$ up to the second order around $\lambda = 0$, we obtain:

$$N(\lambda) = N(0) + N'(0)(\lambda - 0) + \frac{N''(0)(\lambda - 0)^2}{2!} + R_3(\lambda)$$

$$= (1 - 1) + (1 - 1)\lambda + \frac{1}{2}\lambda^2 + R_3(\lambda) =$$

$$= \frac{1}{2}\lambda^2 + R_3(\lambda)$$

Lagrange theorem states that there exists $\xi \in (0, \lambda)$ such that $R_3(\lambda) = -\frac{1}{6}e^{-\xi}\lambda^3$, as a consequence, being $\lambda > 0$, necessarily $R_3(\lambda) < 0$ and the following upper bound holds:

$$N(\lambda) = \frac{1}{2}(\lambda)^2 + R_3(\lambda) < \frac{1}{2}(\lambda)^2$$

Rewriting $\lambda$ as C/A we obtain:

$$N(C/A) = \frac{1}{2}(C/A)^2 + R_3(C/A) < \frac{1}{2}(C/A)^2$$

**Nicola Zingirian**

is Associate Professor with the Department of Information Engineering at the University of Padova since 2002, where he teaches Computer Networks. He is member of the CIPI (Inter-university center in the research of IT platforms). His research interests include Computer Network performance evaluation, Mobile Applications, Sensor Networks. Since 2004 he founded and is directing a spin-off company, Click & Find, that is currently an international market leader in the real-time remote monitoring of fuel road transportation.

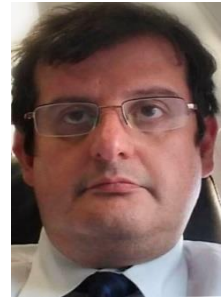

**Michele Benini**

obtained the B.Sc degree in 2017 in Computer Engineering from the University of Padova discussing a thesis in the click spam methods. Now he is M.Sc student in Information Engineering with the same university.

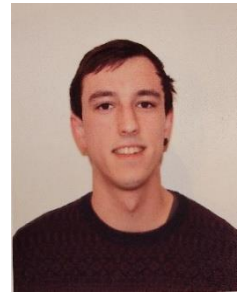